\documentclass[a4paper]{article}

\usepackage{INTERSPEECH2020}
\usepackage{dirtree,siunitx,url}
\usepackage{multirow}
\usepackage{subfigure}
\usepackage{graphics,graphicx}
\usepackage{verbatim}

\newcommand{\Fig}[1]{\textbf{Figure~\ref{fig:#1}}} % refer to figure
\newcommand{\Figure}[3]{\vspace{-0mm} \includegraphics[width=#1,clip]{./figure/#2.pdf} \vspace{-0mm} \caption{#3} \vspace{-0mm} \label{fig:#2}} 

\newcommand{\drawfig}[4]{ % draw figure 
  \begin{figure}[#1]
  \begin{center}
  \Figure{#2}{#3}{#4} 
  \end{center} 
  \end{figure}
}
\newcommand{\drawfigwide}[4]{ % draw figure 
  \begin{figure*}[#1]
  \begin{center}
  \Figure{#2}{#3}{#4} 
  \end{center} 
  \end{figure*}
}

\title{PJS: phoneme-balanced Japanese singing voice corpus}
\name{Junya Koguchi$^{1}$ and Shinnosuke Takamichi$^{2}$}
\address{$^1$ Meiji University, Japan. \\
         $^2$ Graduate School of Information Science and Technology, The University of Tokyo, Japan. }
\email{cs202027@meiji.ac.jp, shinnosuke\_takamichi@ipc.i.u-tokyo.ac.jp}

\begin{document}
\maketitle
%\setlength{\abovedisplayskip}{5pt} % 上部のマージン
%\setlength{\belowdisplayskip}{5pt} % 下部のマージン
%\renewcommand{\subfigcapskip}{5pt} %キャプションと図の間の間隔調整
%\allowdisplaybreaks

\begin{abstract}
    This paper presents a free Japanese singing voice corpus that can be used for highly applicable and reproducible singing voice synthesis research. A singing voice corpus helps develop singing voice synthesis, but existing corpora have two critical problems: data imbalance (singing voice corpora do not guarantee phoneme balance, unlike speaking-voice corpora) and copyright issues (cannot legally share data). As a way to avoid these problems, we constructed a PJS (phoneme-balanced Japanese singing voice) corpus that guarantees phoneme balance and is licensed with CC BY-SA 4.0, and we composed melodies using a phoneme-balanced speaking-voice corpus. This paper describes how we built the corpus.
\end{abstract}

\section{Introduction}
     With the recent developments in deep learning and signal processing, we can now synthesize high-quality singing voices. Various deep learning architectures have been utilized (e.g., feed-forward~\cite{nishimura2016singing}, recurrent~\cite{kim18lstmsingingsynthesis}, and auto-regressive types~\cite{blaauw2017neural}), and many products have been launched (e.g., Sinsy~\cite{sinsy} and NEUTRINO~\cite{neutrino}).
    
    Freely available singing voice corpora contribute to applicable and reproducible singing voice synthesis research. Corpora are being developed in many languages (e.g., Chinese~\cite{hsu10mir1kcorpus}, English~\cite{duan13nuscorpus}, etc.~\cite{sisec16}). The leading Japanese corpus, the large RWC Music Database~\cite{goto05hammingdatabase,goto02rwcdatabase}, was developed 15 years ago. While the RWC corpus was designed for more general use in music information research, the recently developed Tohoku Kiritan database~\cite{kiritan} was designed for singing voice synthesis. The corpus contains a selection of 50 songs made up of children’s songs and anime songs. By comparing these corpora, we aim to develop a smaller corpus for easy-to-train machine learning. The HTS demo~\cite{hts} and JVS-MuSiC~\cite{tamaru20jvs_music} examples never guarantee phoneme balance, which is an important factor in creating a smaller corpus. Phoneme imbalance typically results in phonetic lack in synthesized singing voices.
    
    This paper describes the construction of a phoneme-balanced singing voice corpus named the “phoneme-balanced Japanese singing voice” (\textit{PJS}). Using the Voice Actress Corpus~\cite{seiyu100}, a phoneme-balanced speaking voice corpus, we composed melodies for 100 sentences. Additionally, our corpus contributes the following:
        \begin{itemize} \leftskip -5mm 
        \item[] \textbf{Singing and speaking voices}: We recorded both singing voices and parallel speaking voices. This paired data contributes to speaking-singing research (e.g., \cite{ohishi05}).
        \item[] \textbf{Descriptions of compositions}: We noted descriptions of melody compositions. These descriptions contribute to natural-language-based music information research.
        \item[] \textbf{CC BY-SA 4.0 license}: All the data in our corpus is licensed with CC BY-SA 4.0. Therefore, our corpus is available for both research and commercial use, unlike existing corpora~\cite{hsu10mir1kcorpus,duan13nuscorpus,sisec16,goto05hammingdatabase,goto02rwcdatabase,kiritan,hts}.
         \item[] \textbf{Availability online}: Our corpus can be freely downloaded from our project page~\cite{pjs_url}.
        \end{itemize}

    The following sections describe the details of the corpus.
    
\section{Corpus design}
    \subsection{Directory structure}
        Here, we list the directory structure of our corpus. \textit{[SENTENCE\_ID]} in directory PJS100\_\textit{[SENTENCE\_ID]} is the sentence ID of the original speaking voice corpus~\cite{seiyu100}.
            \dirtree{% 
                .1 \includegraphics[width=0.25cm]{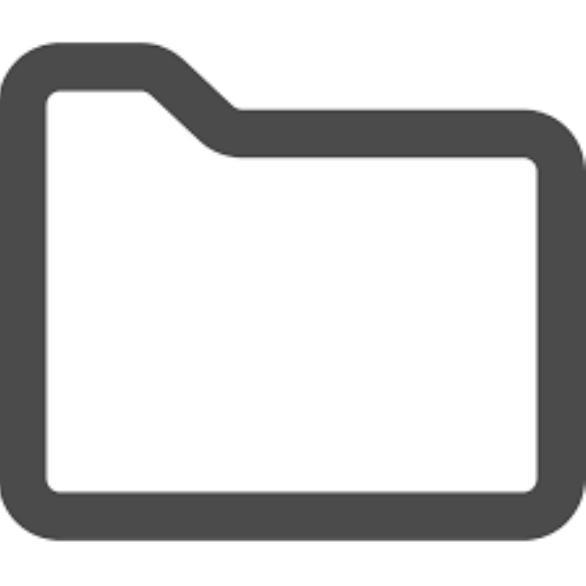} \textbf{PJS100\_001}.
                    .2 \includegraphics[width=0.25cm]{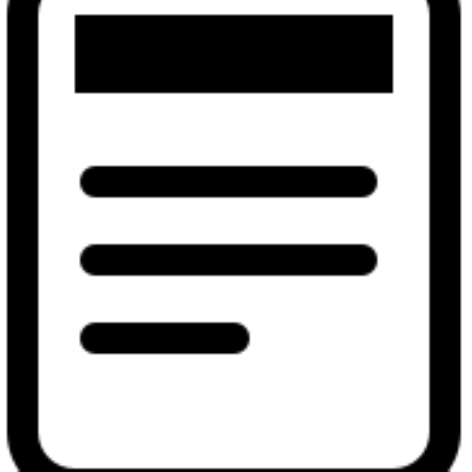} PJS100\_001\_song.wav.
                    .2 \includegraphics[width=0.25cm]{figure/file.pdf} PJS100\_001\_speech.wav.
                    .2 \includegraphics[width=0.25cm]{figure/file.pdf} PJS100\_001.mid.
                    .2 \includegraphics[width=0.25cm]{figure/file.pdf} PJS100\_001.xml.
                    .2 \includegraphics[width=0.25cm]{figure/file.pdf} PJS100\_001.lab.
                    .2 \includegraphics[width=0.25cm]{figure/file.pdf} PJS100\_001.txt.
                .1 \includegraphics[width=0.25cm]{figure/dir.pdf} \textbf{PJS100\_002}.
                .1 \includegraphics[width=0.25cm]{figure/dir.pdf} ....
                .1 \includegraphics[width=0.25cm]{figure/dir.pdf} \textbf{PJS100\_100}.
                }
        The directory PJS100\_\textit{[SENTENCE\_ID]} consists of the following files:
        \begin{itemize} \leftskip -5mm 
            \item PJS100\_\textit{[SENTENCE\_ID]}\_song.wav: singing voice we composed using a sentence from the phoneme-balanced speaking-voice corpus~\cite{seiyu100} as the lyric
            \item PJS100\_\textit{[SENTENCE\_ID]}\_speech.wav: speaking voice that utters a sentence from the phoneme-balanced speaking-voice corpus~\cite{seiyu100}
            \item PJS100\_\textit{[SENTENCE\_ID]}.mid: MIDI file we used as the guide melody during recording
            \item PJS100\_\textit{[SENTENCE\_ID]}.xml: musicXML file that describes musical note information
            \item PJS100\_\textit{[SENTENCE\_ID]}.txt: musical information that songs use (e.g., genre, scale, artist, etc.)
        \end{itemize}

        We composed and recorded 100 phoneme-balanced sentences~\cite{seiyu100}. The following sections describe the composition and recording conditions.
        
    \subsection{Composition conditions}
        A native Japanese male in his twenties composed all the songs. He is not a professional composer but has work experience using his singing, composing, and recording skills. He composed melodies within his range using each of the phoneme-balanced sentences. The musical notes he composed were written in PJS100\_\textit{[SENTENCE\_ID]}.xml. He composed a variety of melodies (based on genre, scale, etc.). Descriptions of the compositions were written in PJS100\_\textit{[SENTENCE\_ID]}.txt. He also made a MIDI file (PJS100\_\textit{[SENTENCE\_ID]}.xml) of the composed melody to guide the recording described below.
    
    \subsection{Recording conditions}
        The composer was also the singer. While listening to the guide melody generated from the MIDI file, he recorded his singing voice so that his pitch and tempo would be as in sync with the guide as possible. To avoid the proximity effect of the microphone, we let him maintain \SI{15}{cm} between the microphone and his mouth. The recording environment was a simple soundproof room in which we attached sound-absorbing materials to the walls. The recording environment was not an anechoic chamber, so we recorded $15$-second background noise each recording day for noise reduction after the recording. We used a Lewitt LCT 441 FLEX (cardioid mode)~\cite{lewitt} microphone, a JZ MICROPHONES Pop Filter~\cite{jz} windscreen, and an RME Fireface UCX~\cite{rme} audio interface.

        We also let him record his speaking voice in the same manner. We saved the singing and speaking voices in the \SI{48}{\kilo\hertz}-sampled, $24$~bit-encoded RIFF WAV format.

\section{Corpus specifications}
    \subsection{Data statistics}
        The data size of the singing voice was larger than that of the speaking voice. The recording of the singing voice was 27.20 minutes long, and the recording of the speaking voice was 12.09 minutes long. Therefore, texts are shared between singing and speaking voices, but the duration of the singing voice is longer than that of speaking voice. This is consistent with existing work~\cite{ohishi05}.

        \Fig{key_hist} and \Fig{tempo_hist} show histograms of the keys and tempos of our corpus, respectively. As \Fig{key_hist} shows, the tonics are well-balanced, while there are fewer songs in minor keys than in major keys. Moreover, as \Fig{tempo_hist} shows, the tempos are distributed in a range between 80 to 160 beats per minute (BPM), indicating that this corpus may be unsuitable for synthesizing songs with extremely slow or fast tempos or in a minor key. 

            % key histogram
            \drawfig{t}{\columnwidth}{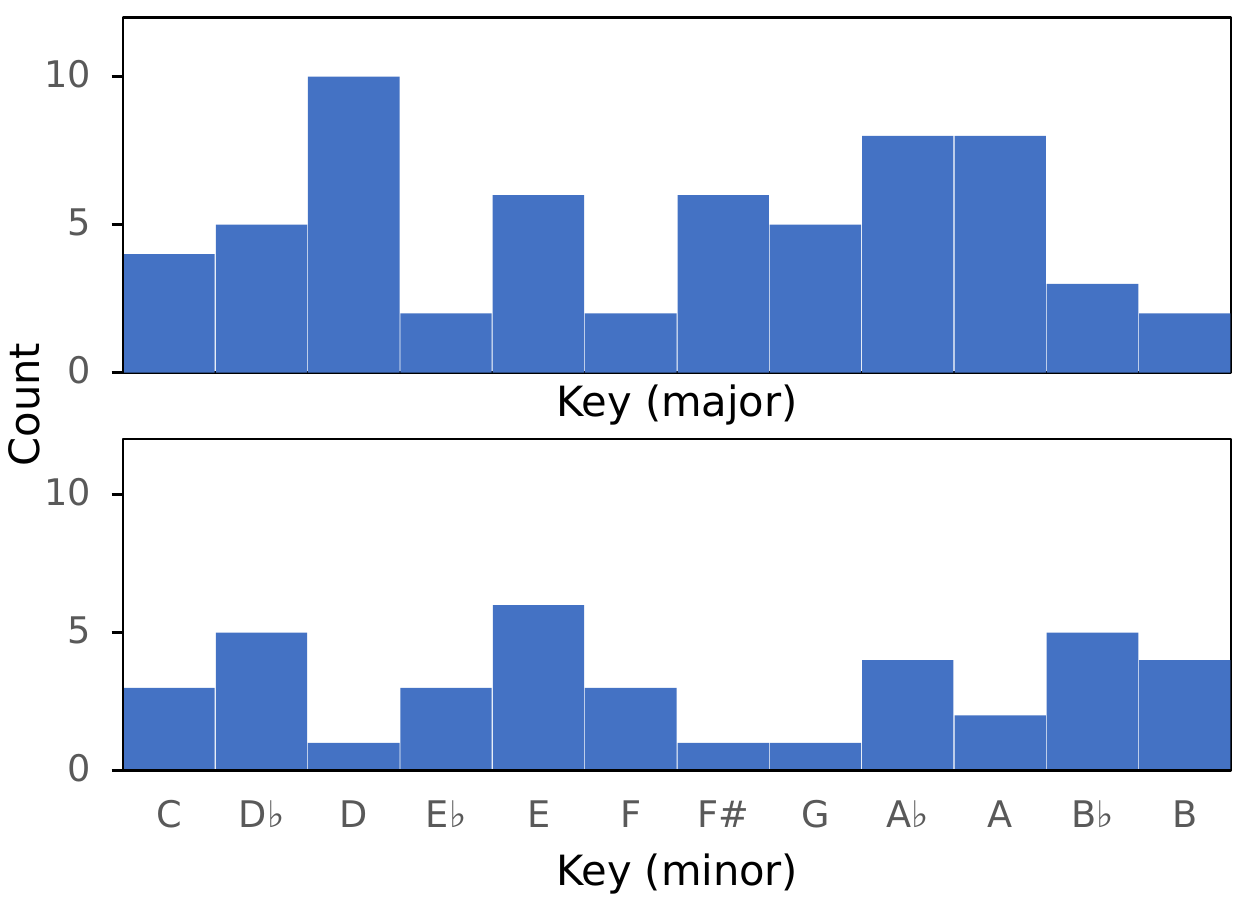}{Key histogram of our corpus. There are fewer songs in minor keys than in major keys.}
    
            % tempo histogram
            \drawfig{t}{\columnwidth}{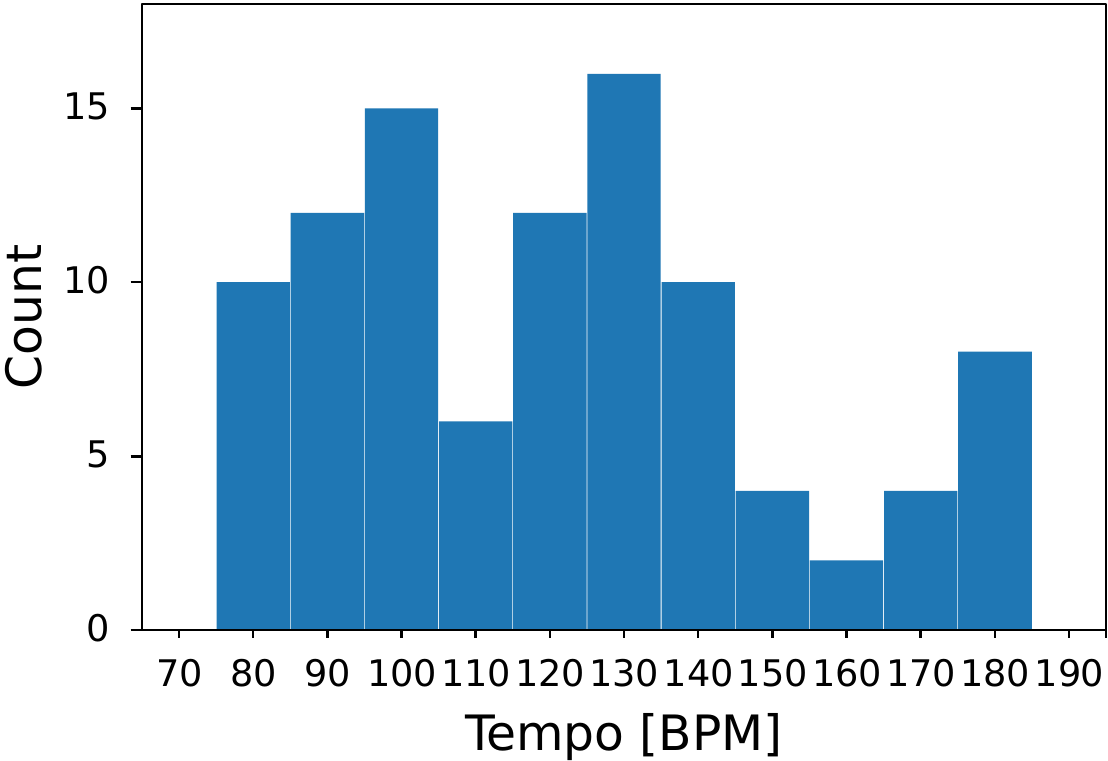}{Tempo histogram of our corpus. The songs only range from 80 to 160 beats per minute (BPM).}

    \subsection{Music score analysis}
        Japanese songs typically use one musical note per Japanese syllable but not always. \Fig{pjs001_trim} is an example of such an exception, PJS100\_001.xml. The multisyllabic notes \textit{to-o}-ji and myo-o-\textit{o-o} can be found on the first and second musical bars, respectively, where ``-'' indicates the syllable boundary.  This means special processes (e.g., copying notes to each syllable~\cite{nakamura14}) are needed to train singing voice synthesizers.
        
            % music score 
            \drawfigwide{t}{2.\columnwidth}{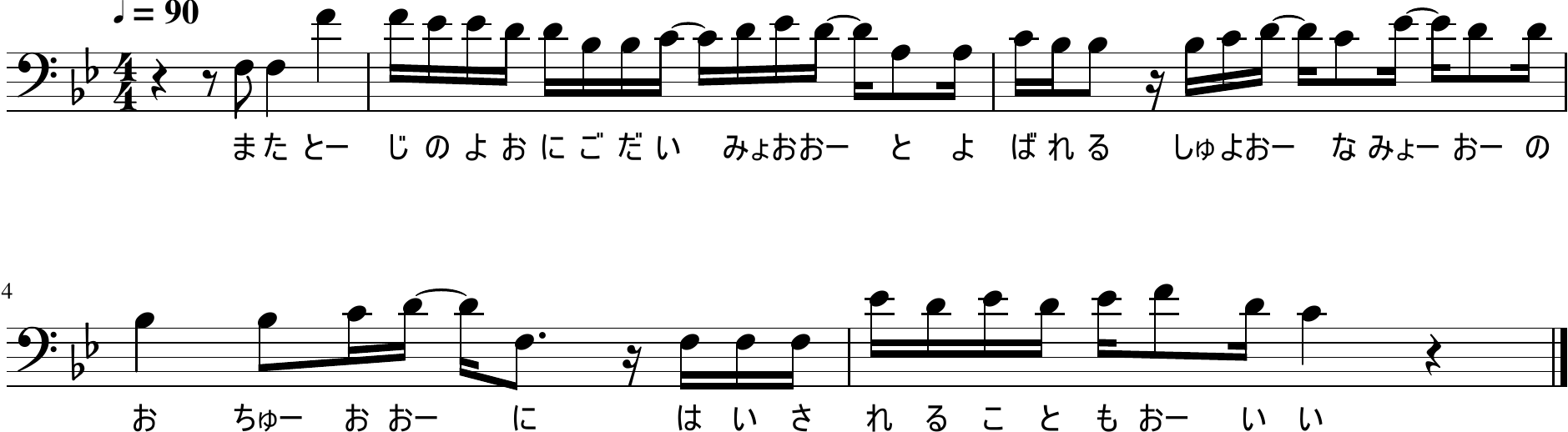}
            {Score of PJS100\_001.xml. The lyrics are ``mata tooji no yoo ni godai myoooo to yobareru shuyoo na myoooo no chuuoo ni haisareru koto mo ooi.'' Most (but not all) individual notes correspond to a single syllable. Some notes correspond to multiple syllables, such as \textit{to-o}-ji on the first musical bar and myo-o-\textit{o-o} on the second musical bar, where ``-'' indicates the syllable boundary.}

\section{Conclusion}
    This paper presented the PJS corpus, a freely available phoneme-balanced Japanese singing voice corpus. We confirmed the phoneme balance in our corpus by composing music based on a phoneme-balanced speaking-voice corpus. Our corpus consists of singing voice data, parallel speaking-voice data, and the musical information that songs use. Therefore, our corpus can contribute to research areas beyond singing voice synthesis. In our future work, we will add a variety of singing styles, such as falsetto and growl voices.

    The PJS corpus is available on our project page~\cite{pjs_url}. All the data is licensed with the CC BY-SA 4.0 license.

\textbf{Acknowledgements:}
    Part of this research was supported by the GAP Foundation Program of the University of Tokyo.
    
\bibliographystyle{IEEEbib}
\bibliography{bib/tts.bib, bib/pjs.bib}

\end{document}